\documentclass[12pt]{article}
\usepackage{graphicx}
\usepackage{epsf}
\newcommand{\be}[1]{\begin{equation}\label{#1}}
\newcommand{\ee}{\end{equation}}
\newcommand{\bea}[1]{\begin{eqnarray}\label{#1}}
\newcommand{\eea}{\end{eqnarray}}
\begin{document}
\begin{center}
{\LARGE \bf  Presynaptic calcium dynamics\\[-1mm]
of neurons in answer to\\[2mm]
various stimulation protocols}\\
\vspace{4mm}
Michael Meyer-Hermann, Frido Erler, Gerhard Soff\\
\vspace{4mm}
Institut f\"ur Theoretische Physik, TU Dresden,
D-01062 Dresden, Germany\\
E-Mail: meyer-hermann@physik.tu-dresden.de\\
\end{center}

\vspace*{4mm}

\noindent{\bf Abstract:}
We present a new model for the dynamics of the presynaptic intracellular
calcium concentration in neurons evoked by various stimulation protocols.
The aim of the model is twofold: We want to discuss the calcium transients
during and after specific stimulation protocols as they are used to
induce long-term-depression and long-term-potentiation. In addition
we would like to provide a general tool which allows the
comparison of different calcium experiments. This may help
to draw conclusions on a wider base in future.

\section*{Introduction}

A most fascinating challenge in neuron biology is a deep 
understanding of the mechanisms involved in long-term-effects (LTE)
such as long-term-potentiation (LTP) and long-term-depression 
(LTD).
The multiplicity of possibly important mechanisms is immense and
is explored in a great variety of experiments \cite{Bau96}. 
However, the interpretations of those experiments are not as
conclusive as they could be if it was possible to better compare
experiments executed on different systems. 
We claim that a lot of detailed information on LTE is hidden 
in presently available experiments.

One possible way to uncover this hidden knowledge is to construct a
tool which is able to translate different 
experiments into a common language and then to compare them
quantitatively. Such a tool is provided here for the
analysis of the intracellular calcium concentration
in presynaptic nerve terminals.
It is very well established by experiments that 
LTE are mostly connected with dynamical changes in the
calcium concentration \cite{Bli93}. 
Larger calcium concentrations induce
a lot of mechanisms that may influence the transmission
efficiency of synapses. This applies not only to the postsynaptic
side of the synapse, where an calcium influx is believed
to be necessary for the induction of LTP \cite{Tsi90}.
Intracellular calcium also induces 
exocytosis on the presynaptic side of the synapse
and therefore is in parts responsible for the release of 
neurotransmitter into the synaptic cleft \cite{Dod67}. 
It is believed that calcium/calmodulin
dependent protein kinases may trigger the amount of
released neurotransmitter. 
This has been shown for example
for the calcium/calmodulin kinase II in the squid giant
synapse \cite{Lli85}.
Therefore, we think that the intracellular calcium concentration
is an appropriate observable to study the induction of 
effects as LTD and LTP from the very beginning.

Already on this early level of LTE-induction a
comparative quantitative evaluation of different 
experiments may lead to new insights.
Corresponding experiments have been done using different systems
or using the same system under different conditions.
In addition, a dependence of the presynaptic calcium dynamics on 
the corresponding target cell has been observed \cite{Koe00}.
In order to compare those experiments quantitatively one has to
determine characteristics of the experiment
that are important for the calcium dynamics
and to introduce them into the model terminology.

In the following we will develop a corresponding model which
is based on previously developed models of the intracellular
calcium dynamics. 
The number of models describing presynaptic 
calcium dynamics is comparably small \cite{Neh92,des93,Hel97,Sin97}.
The main focus of model work in this field
lies on the postsynaptic calcium dynamics
\cite{Neh92,Gam87,Hol90,Jaf94,Sie94,Sch95,des98b,Vol99,Hol00} 
(for a review we refer to \cite{des98}). 
This is surely
related to the fact that dendrites are better accessible
in experiment than presynaptic boutons. Therefore, we use
the knowledge which has been established on the postsynaptic
side of the synapse in order to build a new model which
can be used for comparison of presynaptic calcium dynamics
in different experiments. Presynaptic measurements became
more frequent due to technical developments in the last years
\cite{Sab98}.

The model has to be adjusted to specific experiments
in a well-defined procedure, which is illustrated for
the example of presynaptic nerve terminals in the
rat neocortex \cite{Koe00}. 
The results found here are discussed in the context of LTE.
For more details of the model and the results presented here 
we refer to \cite{Erl02}. This especially concerns a more
detailed analysis of the model assumptions and the robustness
of the results.

\section*{The model on the level of single proteins}

We construct a new deterministic one-compartment model 
for the presynaptic calcium dynamics.
The intracellular calcium concentration is changed in response to
variations of the membrane potential by
an ion-flux through voltage-gated calcium channels.
The flux is driven by the electro-chemical gradient at
the membrane. Intracellular calcium is bound by an
endogenous buffer. Only a small number of ions remains
free and increases the concentration of free calcium.
An increased calcium concentration is reduced by two
types of transport proteins:
ATP-driven calcium pumps and natrium-calcium exchangers.
Their activity basically depends on the free
calcium concentration. A leakage transmembrane 
calcium current is assumed, 
that compensates the activity of
transport proteins and channels in the rest state
of the cell.

The model is formulated in terms of
a set of ordinary coupled differential equations for the 
intracellular calcium concentration $c$:
\be{main}
\frac{dc}{dt}
\;=\;
\frac{G}{zF}\,
\frac{
\rho_{\rm U} J_U\left(g_{\rm U}(U),U(t),\overline{U}(c)\right)
- \rho_{\rm P} J_{\rm P}(g_{\rm P}(c))
- \rho_{\rm E} J_{\rm E}(g_{\rm E}(c))
+ L}{1+\Theta_b(c)+\Theta_i(c)}
\ee
where $\rho_{\rm U, P, E}$ are the surface densities of
the voltage-gated channels (HVA), 
the PMCA-type calcium pumps, and the natrium-calcium
exchanger (type 1), respectively. 
\be{singleJ}
J_{\rm U, P, E}\;=\; g_{\rm U, P, E} \,I_{\rm U, P, E}
\ee
are the corresponding single protein currents. They are determined
by the the single protein open probabilities $g_{\rm U, P, E}$,
where $g_{\rm P, E}$ depend on $c$ according to standard Hill equations
\cite{Zad90}
for the pumps and the exchangers 
with Hill coefficient $2$ \cite{Ell97} and $1$ \cite{Bla99}, 
respectively.  
The open probability for the voltage
gated calcium channels (HVA) obeys \cite{Mag95}
\be{gu_open}
\frac{dg_{\rm U}(U)}{dt}
\;=\; \frac{1}{\tau}
\left\{\left[1+\exp\left\{\frac{U_{1/2}-U}\kappa\right\}\right]^{-1}
-g_{\rm U}(U)\right\}
\quad ,
\ee
with $\tau=1ms$ the channel mean open time, $U_{1/2}=3mV$ the half activation
voltage, and $\kappa=8mV$ the steepness factor. The values are taken
from a representative HVA-channel \cite{Mag95}.
$I_{\rm U}$ is the voltage dependent current which is assumed to follow Ohms law 
with open channel conductivity of $17pS$. The electrochemical gradient
is calculated with respect to $\overline{U}(c)$, the 
(calcium dependent) calcium reversal potential, which depends on $c$
according to the Nernst equation.

$I_{\rm P}=10^{-17} A$ and $I_{\rm E}=40 \,10^{-17} A$ 
are the maximum activities of the transport proteins.
$G$ is a geometry factor (ratio of surface and volume
in the compartment), $z=2$ is the valence of the calcium ions, 
$F$ the Faraday constant,
and $L$ the leak current which is determined by the steady state
condition. 

$U(t)$ is the stimulating transmembrane voltage function. We assume
that the feedback effect of the calcium influx on the membrane potential
is small with respect to standard action potentials. One may think
of a voltage clamped situation. However, for large calcium concentrations
as they may appear in high frequency stimulations, calcium buffers
may saturate and the impact on
the membrane potential may become non-negligible. Such effects are not
considered in the present analysis.

The buffers (endogenous and indicator)
are treated in a quasi-steady state approximation, which
claims that the calcium binds and dissociates faster than the typical
time scale under consideration. Then the dynamical behavior of the
buffers reduces to a correction factor in Eq.~(\ref{main}) which
depends on the calcium concentration only:
\be{buffers}
\Theta_b(c) \;=\; \frac{b_{\rm max} K_b}{\left(K_b+c\right)^2}
\quad \mbox{\rm and} \quad
\Theta_i(c) \;=\; \frac{i_{\rm max} K_i}{\left(K_i+c\right)^2}
\quad .
\ee
Here $b_{\rm max}$ and $i_{\rm max}$ 
are the total concentrations of the 
endogenous buffer and the indicator, respectively. 
$K_{b,i}$ are the
corresponding dissociation constants.
Note that it is important
to include the indicator used in the experiment, because
the indicator is basically acting as additional buffer and
may drastically change the amount of free intracellular calcium.

The differential
equations are solved numerically and the solution describes
the time course of the calcium concentration resulting
in response to single action potentials
or to series of action potentials (as they are used to
induce LTD or LTP). The above mentioned aspired generality
of the model is reflected in a separation of the model
parameters into three classes, described in
the following subsections.

\subsection*{Universality}
{\it The model is universal
enough to be applicable to a wide class
of different neuron types.} To this end the model is based on the
experimental knowledge about {\it single} proteins which are postulated to 
have neuron-type independent properties. These are the single protein
characteristics $g_x$ (including all derived physiological properties)
and the single protein
transmembrane currents $J_x$, where $x$ stands
for the corresponding type of protein (see e.g.~\cite{Ell97,Bla99,Mag95}).

\subsection*{Type specificity}
{\it The model is specific enough to be applicable 
to well defined neuron types.}
This is achieved by the introduction of measurable 
neuron-type specific parameters which has to be determined for each
experiment separately. Basically, these are the protein densities
$\rho_{x}$ in the
membrane. As no space resolution of the calcium concentration is
considered, these densities may be thought as average values over the
whole synaptic membrane. Also the concentration of the endogenous
buffer $b_{\rm max}$ and its dissociation constant $K_b$
belong to the neuron-type specific class of parameters. 
Finally, the surface
to volume ratio $G$ of the synaptic compartment quantitatively
determine the concentration changes due to transmembrane currents.

\subsection*{Condition specificity}
{\it The model includes enough general specifications in order to
adjust the model to specific experimental conditions.}
The form and amplitude of the action potential $U(t)$ is simulated with
a system of coupled differential equations (not shown here) and can
be adapted to the specific action potential used in experiment.
The LTE-stimulation protocols used in
experiment are simulated with a corresponding series of single action
potentials. 
Intracellular calcium concentrations are generally visualized with the help
of calcium indicators.
They act as an additional buffer in the cell and, therefore, may
influence the calcium dynamics.
In the model they are treated in complete
analogy to the endogenous buffer and are characterized by 
the indicator specific dissociation constant $K_i$ and the
used indicator concentration $i_{\rm max}$.

\section*{Adjustment to a specific experiment}

The idea of this semi-universal model for presynaptic calcium dynamics is
to determine the universal parameters using 
single protein experimental data and to maintain the resulting values 
for the evaluation of different systems and experimental conditions. 
{\it Universality} has to be understood as the postulated statement, that
universal parameters are not the parameters that are most sensitive
to a transfer from one experiment to another.
The adjustment of the model to a specific experiment is achieved through
the determination of the system-specific and condition-specific parameters.
Note, that without any exception these are parameters with direct
physiological interpretation. Therefore, most of them may be accessible
up to a sufficient precision in several experiments.

In the described procedure the main part of the model is
determined by sources that are independent of the experiment under
consideration. Therefore, the value of the model is tested by its
capability of reproducing the answer of the calcium concentration
to single action potentials on the basis of those independent
and fixed universal parameters. 
As in general not all {\it specific} parameters
will have been determined in the experiment under consideration, 
we will fit the remaining unknown parameters to the single
action potential calcium response.
With the help of a thus defined model it should be possible
to analyse the measured calcium transients evoked by LTD or LTP
stimulation protocols.

\begin{figure}[ht]
\begin{center}
\hspace*{14mm}
\includegraphics[height=9cm, width=11cm]{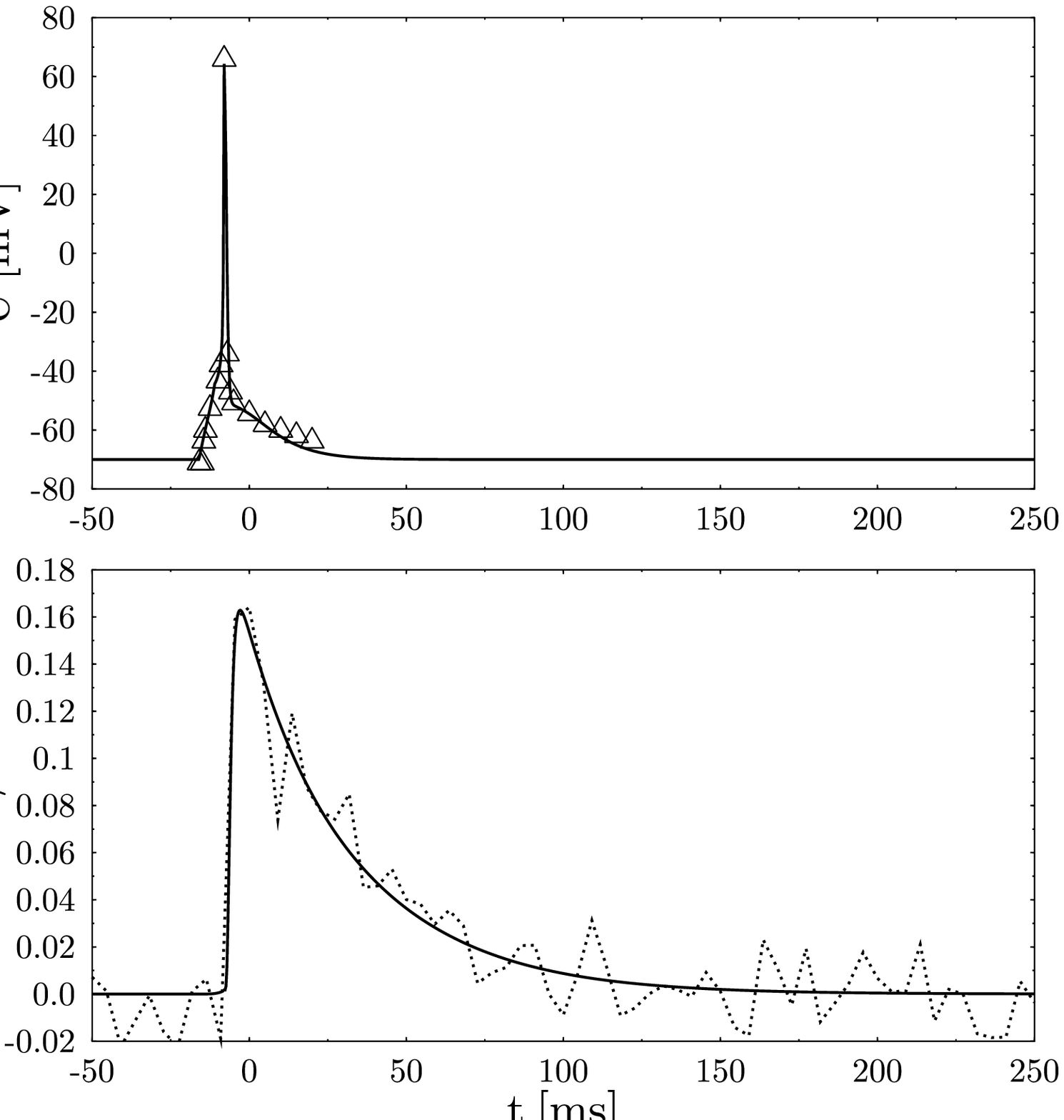}
\end{center}
\caption[]{The calcium transients evoked by single action potentials
in single
boutons of pyramidal neurons in the rat neocortex ($100\mu M$
magnesium green). 
The dotted line shows the experimental values \cite{Koe00}
(we thank H.J. Koester and B. Sakmann for kind permission
to reproduce the data), and the full line
shows the model result with fitted protein densities.}
\label{single}
\end{figure}

This has been executed for an experiment
on single nerve terminals of pyramidal neurons in the neocortex of rats
\cite{Koe00}: Most specific parameters used in the model are
directly accessible in this experiment. This concerns for example
the form of the action potential applied to the nerve terminal, 
the geometry (i.e.\,the surface to volume ratio $G=3/\mu m$),
the concentration of endogenous buffer $b_{\rm max}=120\mu M$, 
and the characteristics
and concentrations of the used calcium indicator 
(magnesium green). Therefore, we are in the situation
that the whole model is a priori determined either by independent sources
(concerning universal parameters) or by the available data from
the experiment under consideration. The only unknown parameters
are the surface densities of the calcium transport proteins.
These are fitted
to the measured calcium transient evoked by single action potentials
($\rho_{\rm U}=2.65/\mu m^2$, $\rho_{\rm P}=4000/\mu m^2$, and
$\rho_{\rm E}=\rho_{\rm P}/30$).
The result Fig.~\ref{single} shows, 
that the measured calcium transient is reproduced correctly. 
%

\section*{LTE stimulation}
\begin{figure}[ht]
\begin{center}
\hspace*{14mm}
\includegraphics[height=9cm,width=11cm]{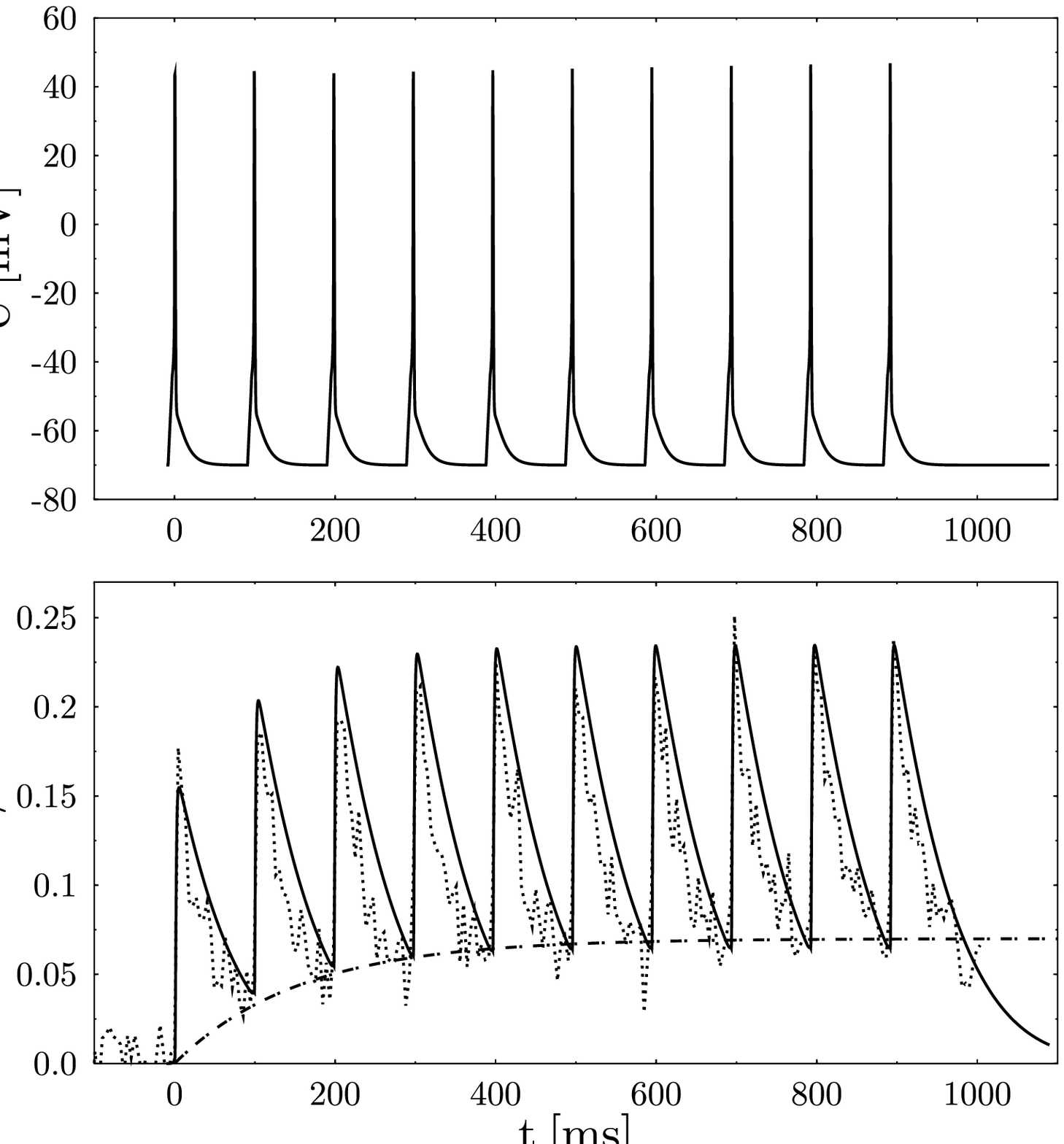}
\end{center}
\caption[]{The calcium transients evoked by a $10 Hz$ stimulus
in single boutons of pyramidal neurons in the rat neocortex
($500\mu M$ magnesium green).
The full line represents the model result and the
dotted line the corresponding measured
transients \cite{Koe00} 
(we thank H.J. Koester and B. Sakmann for kind permission
to reproduce the data).
The dashed line represents the best single exponential fit to
the experimental base line values \cite{Koe00}.}
\label{10hz}
\end{figure}

In order to check if the thus defined model has predictive power,
we calculate the intracellular calcium transients evoked by
series of action potentials with varying frequency.
Basically, the model parameters remain unchanged. 
As in the corresponding experimental setup (see \cite{Koe00} Fig.\,9)
a different action potential (compared to Fig.\,\ref{single}) has been
used, the action potential is adapted in the model and
the channel densities are fitted to the {\it single} action potential
calcium response, correspondingly 
($\rho_{\rm U}=2.77/\mu m^2$, $\rho_{\rm P}=3400/\mu m^2$, and
$\rho_{\rm E}=\rho_{\rm P}/30$). 
In addition the indicator
concentration is increased, which is considerably higher in this experimental
setup. Now the $10\,Hz$ stimulus is applied and the result 
is shown in Fig.~\ref{10hz}. 
%
The model result is in quantitative agreement
with the calcium transients seen in the experiment: The intracellular
calcium concentration reaches a new baseline level during the
stimulation process which breaks down when the stimulus is switched
off. The calcium concentration oscillates on the top of the new base
line in coherence with the stimulation potential. The calcium reducing
processes are slightly to slow in the model.

On this basis we can calculate the calcium transients in
response to stimulation protocols with various frequencies. We find
that calcium transients do not overlap for low frequencies
typical for LTD-induction. The emergence of a new baseline in the calcium
concentration at frequencies around $10\,Hz$ may be interpreted as
threshold for the induction of LTP. Note, that this threshold frequency
strongly depends on the used calcium indicator concentration.
This is especially relevant for the interpretation of experimental
results.
A stimulation with frequencies
around $50\,Hz$ (typical for the induction of LTP) leads to
a more pronounced enhancement of the calcium baseline. This
qualitative behavior is in agreement with experiments
carried out on dendritic spines of pyramidal neurons
\cite{Hel96}.

\vspace*{-2mm}
\section*{Conclusion}

Our new model for transients of the
presynaptic intracellular calcium concentration
evoked by various stimulation protocols 
reproduces the general behavior observed in experiment.
It is exclusively constructed with parameters that have
a direct physiological interpretation. Its basis are
the single proteins properties. The characteristics of
single proteins are considered to be universal in the sense
that they remain unchanged for different experimental setups.
The model has been adjusted to a specific experiment that measured
intracellular calcium transients in
nerve terminals of pyramidal neurons of the rat neocortex.
To this end the parameters specific for this experiment have
been extracted from it or, if not available, have been fitted
to the single action potential calcium response.
The model results turned out to be in quantitative agreement with the
experiment. This applies not only to the
presynaptic calcium concentration response 
to single action potentials but also to $10\,Hz$ stimuli.
We did not find any reason
for the involvement of calcium-induced-calcium-release in
the induction of LTE. 
However, it seems that an additional
mechanism (e.g.~calcium channel inactivation)
may be necessary to understand the induction
of LTD on the level of presynaptic calcium transients.

More generally, the separation of universal and specific 
parameters enables us to analyse different results observed
in several experiments. With the help of the new model
one may decide, if those differences are significant or
due to different experimental setups.

\end{document}